\documentstyle[12pt,epsf]{article}

\begin{document}

\newcommand{\sect}{\section}

\renewcommand {\theequation}{\thesection.\arabic{equation}}
\renewcommand{\sect}[1]{\section{#1}\setcounter{equation}{0} }

\newcommand{\eqr}{\begin{eqnarray}}
\newcommand{\rqe}{\end{eqnarray}}

\newcommand{\eq}{\begin{equation}} 
\newcommand{\qe}{\end{equation}}




\begin{flushright}
HUTP-98/A068\\
hep-th/9809199
\end{flushright}

\vspace{2cm}

\begin{center}

{\bf\Large
Quantum Field Theory of Treasury Bonds} 

\vspace{1.5cm}

Belal E. Baaquie\footnote{Permanent address: Department of Physics, 
National University of Singapore, Kent Ridge Road, Singapore 
091174; e-mail:phybeb@nus.edu.sg} 

\vspace{0.2cm}

{\em
Lyman Laboratory of Physics\\
Harvard University\\
Cambridge, MA 02138, U.S.A.
}
\end{center}

\vspace{1.4cm}

\begin{abstract} 

The Heath-Jarrow-Morton (HJM) formulation of treasury bonds in
terms of forward rates is recast as a problem in path integration.  The
HJM-model is  generalized to the case where all the 
forward rates are allowed to fluctuate independently. The resulting theory
is shown to be a two-dimensional Gaussian 
quantum field theory. The no arbitrage condition is obtained and a
functional integral derivation is given for the price of a futures and an
options contract. 
${         }$
${         }$
\end{abstract}

\thispagestyle{empty}

\newpage

\setcounter{page}{1}

\sect{Introduction}

Stochastic calculus is the most widely used mathematical formalism for
modeling financial instruments \cite{dl}, followed by the use of partial
differential equations \cite{pw}. The Feynman path integral is a formalism
based on functional integration and is 
widely used in theoretical physics to model quantum (random)
phenomenon \cite{ad}; it is also 
ideally suited for studying stochastic processes arising in finance. In
\cite{bs} techniques from physics were applied to the study of finance and
in \cite{b} the problem of the pricing of stock options with stochastic
volatility  was studied using the formalism of path integration.  

In this paper, the path integral approach
is continued into the field of interest rates embodied in the modeling of
Treasury bonds. The complexity of this problem is far greater than that
encountered in the study of stocks and their derivatives; the reason being
that a stock at a given instant in time is described by only one stochastic
variable undergoing random evolution whereas in the case of the interest
rates it is the entire yield curve which is randomly evolving and requires
infinitely many independent variables for its description. The theory of
quantum fields \cite{zj} has been developed precisely to study problems
involving 
infinitely many variables and so we are naturally led to the techniques
of quantum field theory in the study of the interest yield curve.  

Treating all the forward rates as independent random variables has also
been studied in \cite{s1,s2,s3,s4} using the formalism of stochastic
calculus. In this approach a stochastic partial differential equation in
infinitely many variables is written. The approach based on quantum field
theory is in some
sense complimentary to the approach based on stochastic partial
differential equations since the expressions for all  financial instruments
are formally given as a functional integral. One advantage of the approach
based on quantum field theory is that the introduction of non-linearities
as well as stochastic volatility is easily incorporated. 

The HJM-model \cite{hjm}is taken as the starting point of this paper. In
Section 2 the HJM-model  is re-expressed in terms of a path 
integral, and the condition of no arbitrage is re-derived in this
formalism. To make the 
formalism more transparent and accessible to readers not familiar with path
integration, the well-known results for the price of futures 
of zero-coupon bonds as well as the price of a European call option and a
cap for a
zero-coupon bond is derived in Sections 3 and 4 respectively. Another more
important reason for these re-derivations is that the prices of these
derivatives are expressed in a form
which can be directly generalized to the case when we model the evolution
of the forward rates using  quantum field theory. 

In Section 5, the HJM-model is 
generalized to the case with independent fluctuations of all the forward
rates; the theory is then seen to consist of a free 
(Gaussian) two-dimensional quantum field theory.  The generalized model has
a new parameter which determines how 
strongly it deviates from the HJM-model.  The condition of no arbitrage is 
derived for the generalized model.  

In Section 6, the formulae for the prices of futures and options of
zero-coupon bonds are obtained explicitly for the Gaussian quantum field
theory. 

In Section 7 some conclusions are discussed as well as possible future
directions of research.    
        
\sect{Path Integral Formulation of the HJM-model}

{\it Bonds} are financial instruments of debt which are issued by
governments and 
corporations to raise  money from the capital markets \cite{j1,rr}. Bonds
have a 
predetermined (deterministic) cash flow; a Treasury bond is an instrument
for which there is no risk of default in receiving the payments, whereas
for corporate bonds there is in principle such 
a risk.  A Treasury {\it zero-coupon bond} is a risk-free financial
instrument which has a single cash-flow consisting of a
fixed pay-off of say \$1 at some future time T; 
its price at time $t<T$ is denoted by $P(t,T)$, with $P(T,T)=1$. 

A Treasury {\it coupon bond} ${\cal B}(t,T)$ has a series of predetermined
cash-flows which 
consists of coupons worth $c_i$ paid out at increasing times  
$T_i$'s, and with the principal worth $L$ being paid at time $T$. ${\cal
B}(t,T)$ is given in terms of the zero-coupon bonds by \cite{j1}
 
\eq
\label{cb}
{\cal B}(t,T)=\sum_{i=1}^Kc_iP(t,T_i)+LP(t,T)
\qe   

From above we see
that a coupon bond is equivalent to a portfolio of zero-coupon bonds.
Hence, if we model the behaviour of zero-coupon bonds, we
automatically have a model for coupon bonds as well  

Consider the forward rate $f(t,x)$, which stands for the spot (overnight)
interest rate 
at future time $x$ for a contract entered into at time $t<x$. The price of a
zero-coupon bond with the value of \$1 at maturity is given by

\eq
\label{p}
P(t,T)=\exp\{-\int_{t}^{T} dx f(t,x)\}
\qe

Note from its definition, the spot rate for an overnight loan at some
(future) time  $t$ is $r(t)$ and is given by

\eq
\label{r}
r(t)=f(t,t)
\qe

The forward rate is a stochastic variable.  In the K-factor HJM-model 
\cite{hjm,j1,rr} the time
evolution for the forward rates is given by (sum over all repeated index)

\eq
\label{df}
{\partial f\over{\partial t}}(t,x)=\alpha(t,x)+\sigma_i(t,x)W_i(t)
\qe

where $\alpha(t,x)$ is the drift velocity term and $\sigma_i(t,x)$ is the
deterministic volatility for the forward rates. From eqn.(\ref{df}) we have

\eq
\label{f}
f(t,x)=f(t_0,x)+\int_{t_0}^{t} dt'\alpha(t',x) +
\int_{t_0}^{t}dt'\sigma_i(t',x)W_i(t')
\qe

The initial forward rate $f(t_0,x)$ is determined from the market, and so
are the volatility functions $\sigma_i(t,x)$.

Each stochastic variable $W_i(t), i=1,2...K$ is an independent Gaussian
white noise given by  

\eq
E(W_i(t)W_j(t'))=\delta_{ij}\delta(t-t')
\qe

Note that the forward rates $f(t,x)$ are driven by random variables
$W_i(t)$ which 
gives the same random 'shock to all the forward rates; the
volatility function $\sigma(t,x)$ weighs this 'shock' differently for each
time $t$ and each x. It is precisely this
feature which we will generalize later such that $f(t,x)$ is taken to be an
{\it independent} random variable for {\it each} $x$ and {\it each} $t$.

To write the probability measure for $W_i(t)$ we discretize $t=m\epsilon$,
with $m=1,2....M=[{t\over\epsilon}]$, and where $t$ takes values in a
finite interval depending on the problem of interest; then the probability
measure is given by

\eqr
\label{PD}
{\cal
P}[W]&=&\prod_{m=1}^{M}\prod_{i=1}^{K}e^{-{\epsilon\over2}W^2_i(m)}\\
\int dW&=&\prod_{m=1}^{M}\prod_{i=1}^{K}\sqrt{\epsilon\over{2\pi}}
\int_{-\infty}^{+\infty} dW_i(m) 
\rqe              

For notational simplicity we take the limit of $\epsilon \rightarrow
0$; note that for purposes of rigor, the continuum notation is
simply a short-hand  for taking the continuum limit of the
discrete multiple integrals given above.  We have, for $t_1<t<t_2$

\eqr
\label{P}
{\cal P}[W,t_1,t_2]&\rightarrow&e^{S_0}\\
\label{s0}
S_0\equiv S_0[W,t_1,t_2]&=&-{1\over2}\sum_{i=1}^K \int_{t_1}^{t_2}dt
W_i(t)W_i(t)\\ 
\int dW&\rightarrow& \int DW
\rqe

\begin{figure}
\centerline{\epsfxsize=5cm \epsfbox{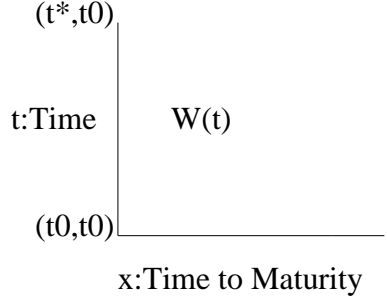}}
\caption{Domain of W(t) }
\label{fighjm}
\end{figure}

The 'action' functional $S_0$ is ultra-local with all the variables being decoupled; generically, $\int DW$
stands for the (path) integration over all the random variables
$W(t)$ which appear in the problem. The integration variables $W(t)$ are
shown in Fig.(\ref{fighjm}).

A path integral approach to the
HJM-model has been
discussed in \cite{cc}; the action 
they derive is different than the one given above since they use a
different set of variables and end up with an action involving the time
derivatives of their variables.  

 A formula that we will repeatedly need
is the generating functional for $W$ given by the path integral 

\eqr
\label{z}
Z[j,t_1,t_2]&=&\int DW e^{\int_{t_1}^{t_2}dt
    j_i(t)W_i(t)}e^{S_0[W,t_1,t_2]}\nonumber  \\
    &=&e^{{1\over2}\int_{t_1}^{t_2}dt j_i(t)j_i(t)}
\rqe

We now derive the no arbitrage condition on the drift velocity $\alpha(t,x)$. 
Recall having a martingale measure is equivalent to having no arbitrage for
the price of the coupon and zero-coupon bonds \cite{h}. The martingale
condition states the 
following: suppose a zero-coupon bond which matures at time
$T$ has a price of $P(t_*,T)$ at time $t_*$ and at time $t_0<t_*$ has a
price $P(t_0,T)$; then the price of the bond at $t_*$,
evolved backward to time $t_0$ and continuously 
discounted by the risk-free spot rate $r(t)$ must be equal to the price
of the bond at time $t_0$.  

In other words, the martingale condition on the zero-coupon bond using
(\ref{f}) is given by 

\begin{figure}
\centerline{\epsfxsize=6.5cm \epsfbox{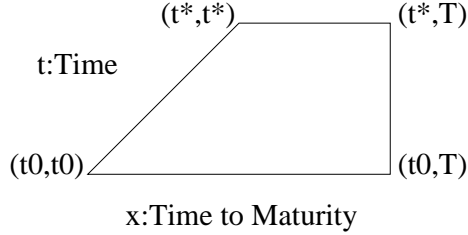}}
\caption{Trapezoidal Domain ${\cal T}$ }
\label{figa}
\end{figure}

\eqr
\label{m}
P(t_0,T)&=&E_{t_0}[e^{-\int_{t_0}^{t_*}r(t)dt} P(t_*,T)] \\
\label{m1}      
  &=& P(t_0,T)e^{-\int_{\cal T}\alpha(t,x)}\int DW
e^{-\int_{\cal T}\sigma_i(t,x)W(t)}e^{S_0[W]} 
\rqe
 
where the {\it trapezoidal} domain ${\cal T}$ is given in Fig.(\ref{figa}) and

\eq
\int_{\cal T}\equiv \int_{t_0}^{t_*}dt
\int_{t}^{T} dx
\qe

\begin{figure}
\centerline{\epsfxsize=5.5cm \epsfbox{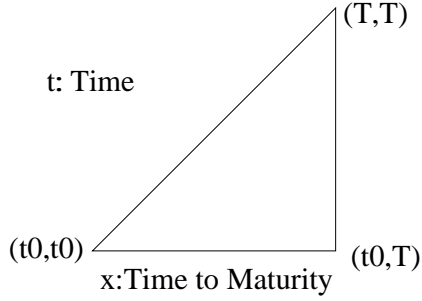}}
\caption{Maximum Domain $\Delta$  }
\label{del}
\end{figure}

We can set $t_*=T$ in above equation; this will change the domain to a
(right isosceles)
{\it triangular} domain $\Delta$  given in Fig.(\ref{del}) and
is the largest domain in the problem. We have, using $P(T,T)=1$, the following

\eqr
P(t_0,T)&=&E_{t_0}[e^{-\int_{t_0}^{t_*}r(t)dt}]\\ 
\Rightarrow e^{\int_{\Delta}\alpha(t,x)}&=&\int DW
e^{-\int_{\Delta}\sigma_i(t,x)W(t)}e^{S_0[W]} 
\rqe

On performing the W-integrations we obtain from (\ref{m1})

\eqr
\label{sm}
e^{-\int_{\cal T}\alpha(t,x)}&=&
e^{{1\over2}\int_{t_0}^{t_*}dt[\int_{t}^{T}dx\sigma(t,x)]^2}   
\rqe   

Dropping the integration
over $t$ we obtain \cite{j1}

\eqr
\label{a2}
\int_t^T\ dx \alpha(t,x)={1\over2}\sum_{i=1}^K[\int_{t}^{T} dx
\sigma_i(t,x)]^2 
\rqe

or equivalently

\eqr
\label{noar}
\alpha(t,x)&=&\sigma_i(t,x)\int_t^x dy\sigma_i(t,y)\\
       &:& Condition for No Arbitrage \nonumber
\rqe

We have see that, as expected, the martingale condition leads to the
well-known no arbitrage condition on the drift velocity of the forward rates.  

Consider the 2-Factor HJM-model with volatilities given by  

\eq
\label{hw}
\sigma_1(t,x)=\sigma_1;\sigma_2(t,x)=\sigma_2 e^{-\lambda (x-t)}
\qe

The no arbitrage condition given in eqn.(\ref{noar}) for this case yields

\eq
\alpha(t,x)=\sigma_1^2(x-t)+{\sigma_2^2\over
\lambda}e^{-\lambda(x-t)}(1-e^{-\lambda(x-t)}) 
\qe

\sect{Futures Pricing in the HJM-Model}

The future and forward contracts on a zero-coupon coupon bond are instruments
that are
traded in the capital markets \cite{j1,rr}.  The forward and future price of
$P(t,T)$, namely $F(t_0,t_*,T)$
and  ${\cal F}(t_0,t_*,T)$ respectively, is the price fixed at time
$t_0<t_*$ for having a zero-coupon bond delivered to the buyer at time $t_*$. 

The difference in the two instruments
is that for a forward contract there is only a single cash flow at the
expiry date of the contract $t_*$. For a futures contract on the other hand
there is a continuous
cash flow from time $t_0$ to $t_*$ such that all
variations in the price of $P(t+dt,T)$ away from $P(t,T)$, for  $t_0<t<t_*$, is
settled continuously between the buyer and the seller, with a final payment
of $P(t_*,T)$ at time $t_*$\cite{j1,rr}. If the
time-evolution of $P(t,T)$ was deterministic, it is easy to see that the
forward and futures price would be equal.    

It can be shown that the price of the futures ${\cal F}$ is given by \cite{j1} 

\eq
{\cal F}(t_0,t_*,T)=E_{t_0}[P(t_*,T)]     
\qe

From eqns.(\ref{f}) and (\ref{P}) we have

\eqr
{\cal F}(t_0,t_*,T)&=&\int DW e^{-\int_{t_*}^{T} dx f(t_*,x)}{\cal
P}[W,t_0,t_*] \\ 
             &=&F(t_0,t_*,T) \exp\Omega_{\cal F}
\rqe

where the forward price for the same contract is given by

\eq
F(t_0,t_*,T)={{P(t_0,T)}\over{P(t_0,t_*)}}
\qe

\begin{figure}
\centerline{\epsfxsize=7.5cm \epsfbox{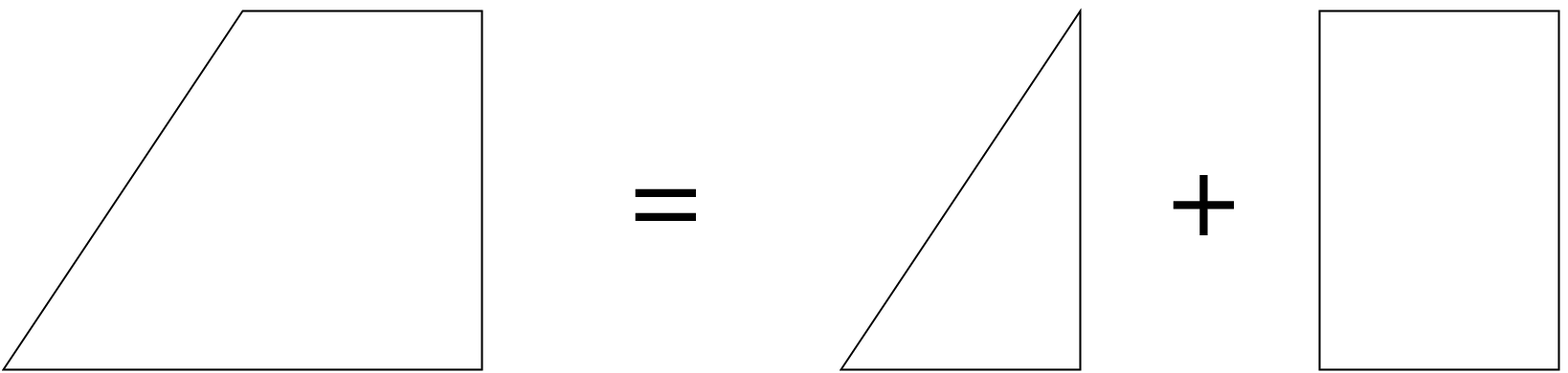}}
\caption{Domain ${\cal T}$=Domain $\Delta_0 \oplus$ Domain ${\cal R}$  }
\label{fac}
\end{figure}

The trapezoidal domain ${\cal T}$ splits into a triangle and a rectangle
shown in Fig.(\ref{fac}) and yields

\eq
{\cal T}=\Delta_0 \oplus {\cal R}
\qe

The futures price is defined over the {\it rectangular} domain ${\cal R}$
given in Fig.(\ref{fac}) and 

\eq
\int_{{\cal R}}\equiv\int_{t_0}^{t_*}dt\int_{t_*}^T dx
\qe

We have 

\eqr
\label{wf}
\exp\Omega_{\cal F}&=& e^{\Omega} e^{-\int_{\cal R}\alpha(t,x)} 
\rqe

with

\eqr
\label{w'}
e^{\Omega}&=&\int DW e^{-\int_{\cal R}\sigma_i(t,x)W_i(t)} e^{S_0} \\
\label{w1}
 &=& \exp{\{{1\over2}\sum_{i=1}^K\int_{t_0}^{t_*}dt [\int_{t_*}^T dx
\sigma_i(t,x)]^2\}}
\rqe  

where eqn. (\ref{w1}) has been obtained by performing the path integration
over the $W$-variables using eqn. (\ref{z}). 

Collecting results from above we have, using the no arbitrage condition
from eqn. (\ref{noar}) and after some simplifications

\eq
\label{w}
\Omega_{\cal F}(t_0,t_*,T)= -\int_{t_0}^{t_*}dt \int_t^{t_*}dx
\sigma_i(t,x)\int_{t_*}^T dx' \sigma_i(t,x') 
\qe

As is expected, the future and forward prices of the zero-coupon bond are
equal if the volatility is zero, that is, the evolution of the zero-coupon
bond is deterministic.
 
Consider the 2-Factor HJM-model with volatilities given in
eqn. (\ref{hw}). Equation (\ref{w}) yields  

\eqr
\Omega_{\cal F}(t_0,t_*,T)&=&-\sigma_1^2(T-t_*)(t_*- t_0)^2 \nonumber\\
                     &-&{\sigma_0^2 \over{2
\lambda^3}}(1-e^{-\lambda(T-t_*)})(1-e^{-\lambda(t_*- t_0)})^2
\rqe

which is the result given in \cite{dl}. 

\sect {Option and Cap Pricing in the HJM-Model}

Suppose we need the price at time $t_0$ of a derivative instrument of a
zero-coupon bond $P(t,T)$ for a contract which
expires at $t_*$ \cite{j1,rr}.  For concreteness  we study  the price of a 
European call option on a zero-coupon bond, namely
$C(t_0,t_*,T,K)$; the  option has a strike price of $K$ and exercise time
at $t_*>t_0$.  

The final value of the option at $t_0=t_*$ is, as required by the contract,
given by 

\eqr
C(t_*,t_*,T,K)&=&(P(t_*,T)-K)_+\\
              &\equiv&(P(t_*,T)-K)\theta(P(t_*,T)-K)
\rqe

where the step function is defined  by

\eq
\theta(x)=\cases {1 & for x$>$0 \cr
                  {1\over2}& for x=0 \cr
                  0 &for x$<$0 \cr}
\qe

For $t_0<t_*$ we have the price of $C$ given by 

\eq
\label{c1}
C(t_0,t_*,T,K)=E_{t_0}[e^{-\int_{t_0}^{t_*}dt f(t,t)}(P(t_*,T)-K)_+]
\qe   

The expectation value in eqn.(\ref{c1}) is  taken by evolving the payoff
function  $(P(t_*,T)-K)_+$
backward to $t_0$, discounted by stochastic spot rate $r(t)=f(t,t)$.

Using the identity

\eq
\label{d}
\delta(z)={1\over {2\pi}}\int_{-\infty}^{+\infty} dp e^{ipz}
\qe

we can rewrite eqn. (\ref{c1}) as

\eqr
\label{c}
C(t_0,t_*,T,K)&=&\int_{-\infty}^{+\infty} dG\Psi(G,t_*,T)(e^G-K)_+
\rqe

where

\eqr
\label{psi} 
\Psi(G,t_*,T)&=&E_{t_0}\Big [e^{-\int_{t_0}^{t_*}dt
f(t,t)}\delta(P(t_*,T)-e^G)\Big ] \\
\label{pint}
       &=&\int_{-\infty}^{+\infty}{dp\over{2\pi}}e^{\Lambda}e^{ip(G+\Lambda_0)}  \\     
\Lambda_0&=&lnF(t_0,t_*,T)
\rqe

Using (\ref{p})and (\ref{f}), we have the following

\eqr
e^\Lambda &=&\int DW e^{-\int_{\Delta_0}\sigma_i(t,x)W_i(t)+ip \int_{{\cal R}}\sigma_i(t,x)W_i(t)}e^{S_0}\over{e^{-\int_{\Delta_0} \alpha(t,x) 
 + ip\int_{{\cal R}}\alpha(t,x)}}
\rqe

Note the interplay of the subdomains $\Delta_0$ and ${\cal R}$ in
determining the price of the option.  Using (\ref{z}) to perform the
integrations over $W$ yields, after 
considerable simplifications and using the no arbitrage condition given by
eqn. (\ref{a2}) the following   

\eq
\label{wop}
\Lambda=-{q^2\over2}(p^2+ip)
\qe

with

\eq
\label{q2}
q^2=\sum_{i=1}^K\int_{t_0}^{t_*}dt[\int_{t_*}^T dx\sigma_i(t,x)]^2
\qe

To  obtain eqn.(\ref{wop}) we have  used the identity

\eq
\int_{t_0}^{t_*} dt[\int_{t_*}^T dx\alpha(t,x)-\int_t^{t_*}dx
\sigma_i(t,x)\int_{t_*}^T 
dy\sigma_i(t,y)]={1\over2} q^2
\qe
  
Performing the Gaussian integration in eqn.(\ref{pint}) yields

\eq
\label{psig}
\Psi(G,t_*,T)=\sqrt{{1\over{2\pi q^2}}}\exp-{1\over{2q^2}}\{G+\int_{t_*}^T dx
f(t_0,x)-{q^2\over2}\}^2 
\qe

Hence from above and (\ref{c}) we recover the well-known result
\cite{j2,jm2} that the 
European option on a zero-coupon has a Black-Scholes like formula with
volatility given by $q$.

For the 2-Factor HJM-model given in eqn.(\ref{hw}) we have

\eqr
\label{q}
q^2&=&\sigma_1^2(T-t_*)^2(t_*-t_0)\nonumber\\
&+&{\sigma_2^2\over{2\lambda^3}}(1-e^{-\lambda(T-
t_*)})^2(1-e^{-2\lambda(t_*-t_0)})
\rqe

A {\bf cap} is a financial instrument for reducing ones exposure to
interest rate fluctuations, and guarantees a maximum interest rate for
borrowings over a  fixed time. A {\bf cap} fixes the maximum interest
for a fixed period from $t_*$ to $t_*+T$; the buyer of the instrument then pays
for this period the maximum of the Libor (London interbank offered rate)
$L\equiv L(t_*,t_*+T)$ {\it or} the cap rate $K$. The cap is exercised at time
$t_*$ and the payments are made, in arrears, at time $t_*+T$. Let the
principal amount be $V$; the value of the cap at time $t_*$ is
then given by 

\eq
Cap(t_*,t_*,T)=VT(L-K)_+
\qe

We have in terms of the forward rates \cite{s2}

\eqr
TL(t_*,t_*+T)&=&e^{\int_{t_*}^{T+t_*}dxf(t_*,x)}-1\\ 
             &=&{1 \over P(t_*,t_*+T)} -1
\rqe

The price of the cap at time $ t_0<t_*$ is given by

\eqr
\label{cap1}
Cap(t_0,t_*,T,X)&=&VE_{t_0}[e^{-\int_{t_0}^{t_*}dt f(t,t)}({1 \over
P(t_*,t_*+T)} -1-TK)_+]\nonumber\\
\\
&=&V\int_{-\infty}^{+\infty} dH\Gamma(H)(e^H-1-TK)_+ 
\rqe   

Carrying out an analysis similar to the one done for the pricing of the
European call option we
obtain, as in (\ref{psig}) (note the minus sign of H)  

\eqr
\Gamma(H)&=&\Psi(-H,t_*,T+t_*)\\
\label{gamma}
       &=&\sqrt{{1\over{2\pi
q_{cap}^2}}}\exp-{1\over{2q_{cap}^2}}(-H+\int_{t_*}^{T+t_*} dx 
f(t_0,x)-{q_{cap}^2\over2})^2  
\rqe

with $q_{cap}$ for the Two-Factor model given similar to (\ref{q}) by  

\eqr
q_{cap}^2 &=&\sigma_1^2 T^2(t_*-t_0)\nonumber\\
&+&{\sigma_2^2\over{2\lambda^3}}(1-e^{-\lambda
T})^2(1-e^{-2\lambda(t_*-t_0)}) 
\rqe

The formula above shows that a cap is equivalent to a European put
option on the zero-coupon bond $P(t_*,t_*+T)$. For a {\it caplet} the time
$T$ is taken to be small so that $L(t_*,t_*+T)\simeq f(t_*,t_*)$, and in
eqn.(\ref{gamma}) we have

\eqr
\int_{t_*}^{T+t_*} dx f(t_0,x)&\simeq& T f(t_0,t_*)\\
q_{cap}^2 &\simeq&T^2[\sigma_1^2(t_*-t_0)
+{\sigma_2^2\over{2\lambda}}(1-e^{-2\lambda(t_*-t_0)})]
\rqe  

\sect{Quantum Field Theory of Treasury Bonds}

As mentioned earlier, in the HJM-model  the fluctuations in the forward
rates at a given time $t$ are given by 
'shocks' which are delivered to the whole curve $f(t,x)$ by a single random
variable 
$W(t)$; clearly, a more general evolution of the instantaneous forward rate
would be to let the whole curve evolve randomly, that is let all the
forward rates fluctuate independently. The only constraint on the random
evolution is that for every instant there be {\it no arbitrage} for the forward
rates.

Since there are infinitely many forward rates, we need an infinite
number of independent variables to describe the
random evolution of the yield curve.  As mentioned earlier the generic quantity
describing such a 
system is a quantum field \cite{zj}; for modeling Treasury Bonds we
will need to study a two-dimensional quantum field in a finite Euclidean
domain. 

We consider the forward rates to be a {\it quantum field}; $f(t,x)$ is
taken to be an
{\it independent} random variable for {\it each} $x$ and {\it each} $t$. 
For notational simplicity we  keep both $t$ and  $x$ 
continuous; in Appendix A, the case with both $t$ and $x$
discrete is analyzed and the continuum limit discussed in some detail. 

For the sake of concreteness, consider the price of an 
options contract (at time $t_0$) of a $T$-maturity zero-coupon bond ; let
the contract 
expire at time $t_*$.  Since the all the forward rates are for the future,
we always have $x>t$; 
hence the quantum field $f(t,x)$ is defined on the trapezoidal domain
${\cal T}$ bounded by
$t=t_0$, $t=x$, $t=t_*$ and $x=T$, as shown in Fig (\ref{figa}). 

We introduce a new parameter $\mu$
which quantifies the strength of the fluctuations in the time-to-maturity
direction $x$; we expect that in the limit of 
$\mu \rightarrow 0$, we should recover the HJM-model.  The simplest term
which can control the fluctuations in the $x$-direction is the gradient of
$f(t,x)$ with respect to $x$. The action given in eqn.(\ref{s0}) is generalized to 

\eqr
\label{sf}
S[f]&=&\int_{\cal T}{\cal L}[f]\\
\label{lf}
{\cal L}[f]&=&-{1\over{2(T-t)}}\Big [\Big \{{{{{\partial
f(t,x)}\over{\partial t}}- \alpha(t,x)}\over{\sigma(t,x)}}\Big\}^2
+ {1\over{\mu^2}}\Big \{ {\partial \over {\partial x}}\Big ({{{{\partial
f(t,x)}\over{\partial t}}- \alpha(t,x)}\over{\sigma(t,x)}}\Big
)\Big\}^2\Big]\nonumber\\ 
\rqe

The initial condition is given by

\eq
t=t_0,t_0<x<T:f(t_0,x):specified
\qe

and the field values of $f(t,x)$ on the rest of the  boundary points of the
trapezoid ${\cal T}$ are arbitrary and are integration variables.
The quantum field theory is defined by integrating over all configurations
of $f(t,x)$ and yields

\eqr
\label{zff}
Z&=&\int Df e^{S[f]}\\
\int Df&=&\prod_{(t,x)\epsilon {\cal T}}\int_{-\infty}^{+\infty}df(t,x)
\rqe

Note that $e^{S[f]}$/ Z is the
probability for different field configurations to occur when the functional
integral over $f(t,x)$ is performed.  

The presence of the second term
in the action given in (\ref{sf}) seems to be justified from the
phenomenology of the forward rates \cite{b2} and is not ruled out by no
arbitrage . This term in the action also implies that all the random 
configurations of $f(t,x)$ which 
appear in the  path integral (\ref{zff})  are continuous functions of
$x$. Forward
rates that are usually observed in the market are continuous \cite{s1}.  

However the random configurations for the
forward rates are nowhere differentiable.  It is noted in \cite{s1} that
there is no evidence to indicate whether the  actual forward rates are
differentiable or not.

The action
given above is suitable for studying formal properties of the forward
rates. However it is often simpler for computational 
purposes to change variables. 
Let $A(t,x)$  be a two-dimensional quantum field; we use the HJM-change
of variables to express $A(t,x)$ in terms of the forward rates $f(t,x)$,
namely  

\eq
\label{dfA}
{\partial f\over{\partial t}}(t,x)=\alpha(t,x)+\sigma(t,x)A(t,x)
\qe

The Jacobian of the above transformation is a constant and hence can be
ignored. The action can now be written as

\eqr
\label{sa}
S[A]&=&-{1\over2}\int_{t_0}^{t_*}dt{1\over{T-t}}\int_t^T
dx\{A^2(t,x)+{1\over\mu^2}({\partial A(t,x)\over{\partial
x}})^2 \}\\
&=&\int_{\cal T}{\cal L}[A] 
\rqe

The quantum field theory is defined by a functional integral over all
variables $A(t,x)$; the values of $A(t,x)$ on the boundary of ${\cal T}$
are arbitrary and are integration variables; this yields the partition
function  

\eq
\label{zaa}
Z=\int DA e^{S[A]}
\qe

Note that eqns. (\ref{dfA}) and (\ref{sa}) can easily be generalized to the
K-Factor case.  It is shown in eqn.(\ref{hjm}) that if we define

\eq
\label{wa}
W(t)={1\over{T-t}}\int_t^TdxA(t,x)
\qe

then for $\mu \rightarrow 0$ we have

\eqr
S[A]&\rightarrow& S_0= -{1\over2}\int_{t_0}^{t_*}dt W^2(t)\\
\int DA &\rightarrow& \int DW
\rqe

From eqn.(\ref{s0}) and above we see that we recover the HJM-model in
the $\mu \rightarrow 0$ limit.  We see from eqn.(\ref{wa}) that the
HJM-model is a drastic 
truncation of the full field theory and only considers the fluctuations of
the  average value of the quantum field $A(t,x)$; it in effect
'freezes-out' all the 
other fluctuations of $A(t,x)$. 

If one thinks of the field $A(t_0,x)$ at some instant
$t_0$ as giving the position of a 'string' \cite{s1,s2}, then in the
HJM-model this string 
is taken to be a {\it rigid} string. The action $S[A]$ given
in (\ref{sa}) allows {\it all} the degrees of freedom of
the field $A(t_0,x)$ to fluctuate independently and can be thought of as a
'string' with string tension equal to ${1\over \mu^2}$; in this
language the HJM-model considers the interest yield curve to be a string
with infinite tension and hence rigid.  

The moment generating functional for the quantum field theory is given by
the Feynman path integral as

\eqr
Z[J]={1\over Z}\int DA e^{\int_{t_0}^{t_*}dt\int_t^TdxJ(t,x)A(t,x)}e^{S[A]}
\rqe

We evaluate $Z[J]$ exactly in Appendix B, and from eqn.(\ref{zf})

\eq
\label{za}
Z[J]=\exp
{1\over2}\int_{t_0}^{t_*}dt\int_t^Tdxdx'J(t,x)D(x,x';t,T)J(t,x') 
\qe

where the propagator $D(x,x';t,T)$ is given from eqn.(\ref{pro}), for 
$\beta=T-t$, by

\eqr
D(x,x';t,T)&=&{{\mu\beta}\over{
\sinh^3(\mu\beta)}}\Big[\sinh\mu(T-x)\sinh\mu(x'-t)\{1+\sinh^2(\mu
\beta)\theta(x-x')\}   
\nonumber\\    
&+&\sinh\mu(T-x')\sinh\mu(x-t)\{1+\sinh^2(\mu\beta)\theta(x'-x)\}\nonumber \\
    &+&\cosh(\mu\beta)\{\sinh\mu(x-t)\sinh\mu(x'-t)\nonumber\\
         &+&\sinh\mu(T-x)\sinh\mu(T-x')\}\Big] 
\rqe           

To understand the significance of the propagator $D(x,x';t,T)$ note that the
correlator of the field $A(t,x)$, for $t_0<t,t'<t_*<T$, is given by

\eqr
\label{aa}
E(A(t,x)A(t',x'))&=&{1\over Z}\int DA e^{S[A]} A(t,x)A(t',x')\\
                 &=&\delta(t-t')D(x,x';t,T)
\rqe

In other words, $D(x,x';t,T)$ is a measure of the effect of a value of
field $A(t,x)$ at maturity $x$ on its value at another maturity $x'$.   

Since $D(x,x';t,T)$ looks fairly complicated, we examine it in a few extreme
limits. In the limit of $\mu \rightarrow 0$ we have 

\eq
\label{d0}
D(x,x';t,T)= 1+O(\mu^2)
\qe

We see that, as expected, all the fluctuations in the $x$ direction are
exactly correlated; in other words the values of $A(t,x)$ for different
maturities are all the same. Defining

\eq
j(t)=\int_t^TdxJ(t,x)
\qe

we have from eqns. (\ref{za}) and (\ref{d0}) that

\eq
Z[j]=\exp {1\over2}\int_{t_0}^{t_*}dtj^2(t)  
\qe

which is the result obtained earlier in eqn.(\ref{z}).

For $\mu \rightarrow \infty$ we have

\eq
D(x,x';t,T)\simeq {1\over2}\mu\beta e^{-\mu|x-x'|} 
\qe

The propagator above has  a simple interpretation; if the field $A(t,x)$
has some value at point $x$, then the field at 'distances'
$x-\mu^{-1}<x'<x+\mu^{-1}$ will tend to have the same value, whereas
for other values of $x'$ the field will have arbitrary values. Hence we
see in this limit that the fluctuations in the time-to-maturity $x$
direction are  
strongly correlated within maturity time $\mu^{-1}$, which is the
{\it correlation time} of the forward rates.

We now  derive the no-arbitrage condition for the action $S[A]$.
Eqn.(\ref{m}) for the martingale is unchanged; generalizing
eqns. (\ref{m1}) and (\ref{sm}) we have

\eqr
\label{no}
\exp\int_{{\cal T}}\alpha(t,x)&=&
{1\over Z}\int DA e^{ -\int_{{\cal T}}\sigma(t,x)A(t,x)}e^{\int_{\cal
T}{\cal L}[A]}  \\
        &=&
\exp{1\over2}\int_{t_0}^{t_*}dt\int_{t}^{T}dxdx'\sigma(t,x)D(x,x';t,T)\sigma(t,
x')\nonumber\\   
 \rqe   

Hence we have

\eq
\label{ad}
\int_{t}^{T}dx\alpha(t,x)={1\over2}\int_{t}^{T}dxdx'\sigma(t,x)D(x,x';t,T)
\sigma(t,x')   
\qe

which is the generalization of eqn.(\ref{a2}), and that of
eqn. (\ref{noar}) is given by 

\eqr
\label{noara}
\alpha(t,T)&=&\sigma(t,T)\int_{t}^{T}dx'D(T,x';t,T)\sigma(t,x')\nonumber\\
&+&{1\over2}\int_{t}^{T}dxdx'\sigma(t,x){{\partial
D(x,x';t,T)}\over{\partial T}}\sigma(t,x') 
\rqe

From the empirical study of forward rate curves, there is evidence \cite{b2}
that the 
naive HJM-model no arbitrage for the drift term $\alpha(t,x)$ is not
adequate since it is quadratic in the volatility; in \cite{b2} an
additional term is added which reflects the market price of risk. In the
approach of field theory, the additional term involving the derivative of
the propagator could provide a better model of no arbitrage for the drift
term.     

For $\mu\to \infty$ we have  

\eq
\alpha(t,T)={1\over2}(T-t)\sigma^2(t,T)+{1\over2}\int_{t}^{T}dx\sigma^2(t,x)
\qe

Note the expression for $\alpha(t,x)$ given above is quite dissimilar from that
of the HJM-model given in eqn.(\ref{noar}), which is the case for
$\mu=0$; the values of $\alpha(t,x)$ given in (\ref{noara}) for $\mu\ne0$
continuously interpolate between the extreme values of $\mu=0$ and
$\mu=\infty$.    

For the Two-Factor case given in eqn.(\ref{hw}), we can exactly solve 
for $\alpha(t,x)$ in terms of the volatilities; the expressions are long and
cumbersome.  For the case of the One-Factor model with only
$\sigma_1\not=0$, we have the exact result that 

\eq
\alpha(t,x)=\sigma_1^2(x-t) 
\qe

which is independent of $\mu$ and the same as the HJM-model.  This result
can be seen directly from the functional integral; since $\sigma_1=$constant,
in the no arbitrage eqn. (\ref{no}) for $\alpha(t,x)$,
we see that it 
only couples to $\int_t^TdxA(t,x)=(T-t)W(t)$; a change of variables then
shows that 
$\alpha(t,x)$ does not couple to $\mu$, and hence the simple result. As we will
see later, the One-Factor model has non-trivial dependence on $\mu$ for
other quantities such as futures and options.
 
We have from eqns.(\ref{dfA}), (\ref{noara})

\eqr
\label{fa}
f(t,x)&=&f(t_0,x)+{1\over2}\int_{t_0}^t
dt'\int_{t'}^{x}dydy'\sigma(t',y){{\partial 
D(y,y';t',x)}\over{\partial x}}\sigma(t',y')\nonumber\\
  &+&\int_{t_0}^{t} dt'\sigma(t',x)[
   \int_{t'}^{x}dy'D(x,y')\sigma_i(t',y')+A(t',x)]
\rqe

\sect{Futures and Option Pricing}

We derive the futures and options pricing using quantum field theory. For
the Two-Factor model all the expressions can be obtained exactly; the
results for the 
$\mu=0$ limit are the same as the HJM-model; we will explicitly give the
results only for the case of $\mu\to\infty$ because the expressions for
general $\mu$ don't add much to ones understanding.   

Equation (\ref{wf}) for the futures price ${\cal F}$  only changes for
$\Omega$; from eqn.(\ref{w'}) we have (note different domains ${\cal R}$
and ${\cal T}$ below)

\eqr
\label{wa'}
e^{\Omega}&=&{1\over Z}\int DA e^{-\int_{\cal R} dx
\sigma(t,x)A(t,x)} e^{\int_{\cal T}{\cal L}[A]} \\
\label{wa1}
 &=& \exp{\{{1\over2}\int_{t_0}^{t_*}dt \int_{t_*}^T dxdx'
\sigma(t,x)D(x,x';t,T)\sigma(t,x')\}}
\rqe  

and, using the no arbitrage condition (\ref{noara}) we obtain the
generalization of (\ref{w}) given by

\eq
\Omega_{\cal F}(t_0,t_*,T)= -\int_{t_0}^{t_*}dt \int_t^{t_*}dx
\sigma_i(t,x)\int_{t_*}^T dx'D(x,x';t,T) \sigma_i(t,x') 
\qe

For the One-Factor Model with only $\sigma_1\ne 0$ we have

\eqr
\lim_{\mu\to \infty}\Omega_{\cal F}(t_0,t_*,T)=&-&{\sigma_1^2\over{4\mu}} 
(t_*-T)(2T-t_0-t_*)+O({1\over \mu^2})
\rqe              

For the price of a European call option $C$, a calculation similar to the
one carried out in 
Section 4 gives the same formula for $\Psi(G)$ given in eqn.(\ref{psig}) with
$q^2$ given in eqn.(\ref{q2}) replaced  by

\eq
\label{qa2}
q^2=\int_{t_0}^{t_*}dt\int_{t_*}^T dxdx'\sigma_i(t,x)D(x,x';t,T)\sigma_i(t,x')
\qe

We have

\eq
\lim_{\mu\to\infty}q^2=\int_{t_0}^{t_*}dt(T-t)\int_{t_*}^T
dx\sigma_i^2(x,t)
\qe

For the Two-Factor model we have

\eqr
\lim_{\mu\to\infty}q^2&=&{\sigma_1^2\over
2}(T-t_*)(t_*-t_0)(2T-t_0-t_*)\nonumber\\ 
&+&{\sigma_2^2\over{8\lambda^3}}\Big[\{1+2\lambda(T-t_0)\}(1-e^{-2\lambda(t_*-t_0)})    
 \nonumber\\     
&-&2\lambda(t_*-t_0)(1-e^{-2\lambda(T-t_*)})\Big]
\rqe

Note for both the futures and option prices, the presence of $\mu$ is like
adding another factor to the model. However, the dependence of the
derivatives on $\mu$ is quite different from that on 
$\lambda$; for instance the no arbitrage condition changes significantly as
$\mu$ goes from small to large whereas no  such effect happens in the case
of $\lambda$; the prices of the derivatives also show non-trivial
dependence on $\mu$.              

If we are interested in pricing any path dependent option or other
derivatives, it is not sufficient to know only the propagator
$D(x,x';t,T)$; the full structure of the action $S[A]$ is then required.  

For example the payoff function of an Asian option at time $t_0$ on a
zero-coupon bond $P(t,T)$ with exercise time $t_*$ is
given by 

\eq
g[P(*,T)]=({1\over{t_*-t_0}}\int_{t_0}^{t_*}dtP(t,T)-K)_+
\qe

Another example is the price of a European call option on a
coupon bond ${\cal B}(t,T)$ given in (\ref{cb}); the payoff function is then   

\eq
g[{\cal B}]= ({\cal B}(t_*,T)-K)_+
\qe

The payoff function $g[A]$ in both the cases above is path
dependent. Expressing all the zero-coupon bonds in terms of the quantum
field $A(t,x)$, the prices of such path dependent options at time $t_0$ are
given by   

\eq
C(t_0,t_*,T,X)={1\over Z}\int DA e^{-\int_{t_0}^{t_*} dt r(t)}g[A]e^{S[A]}
\qe

The computation above can only be performed numerically \cite{he}; for this the
functional integral over $A(t,x)$ has to be discretized, and which is briefly
discussed in Appendix A.

\sect {Conclusions}

We have re-formulated the theory of Treasury bonds in terms of path
integration. The HJM-model has a simple path integral with an ultra-local
action.  The statements about martingale conditions and the evaluation of
futures and options were shown to be calculable in a straightforward manner
using path integration.  The motivation for re-deriving the well-known
results of the HJM-model was firstly to understand the path integral
formulation of the quantities of interest in finance, and secondly, to then
generalize these quantities to the case of quantum field theory.

The quantum field theory of Treasury bonds is more general than the
HJM-model; in particular, the correlation of fluctuations of the forward
rates can be easily modeled to be finite in the field
theory  whereas in the HJM-model {\it all} the fluctuations are exactly
correlated . From the point of view of finance, it is unreasonable to
assume that the all forward rates fluctuate identically as in the
HJM-model; the multi-factors in  HJM-model try and capture the
finite correlation in the time-to-maturity that should  exist for the
forward rates.  

We considered a Gaussian model for the field theory generalization of the
HJM-model as this is the simplest extension, and also because the 
model could be solved exactly. In particular, the formulae for the futures, cap
and option price of Treasury bonds were derived and involved nontrivial
correlations in the volatility of the model.    

We can generalize the  model to account for stochastic volatility of the
forward rates.  This entails introducing another quantum field for modeling
the fluctuations of volatility, and is similar to the quantum mechanical
treatment of volatility for a single security \cite{b}. Stochastic
volatility makes the system highly nonlinear and is treated in
some detail in \cite{bb}.     

The best way of modeling Treasury bonds in practice is a
computational and empirical question \cite{kcc,fl}; only if the field
theory model can be easily calibrated and yields more efficient algorithms,
will it 
it be taken seriously by the practitioners of finance. For the more
theoretical side of finance, the methodology of field theory certainly
adds to the ways of studying and understanding the stochastic processes
which drive the capital markets.  

\vskip 0.3cm

\begin{center}
{\bf Acknowledgments}

I am deeply grateful to Lawrence Ma for many useful and instructive
discussions; most of my 
interest and understanding of this subject is a result of these
discussions. I would like to thank Toh Choon Peng, Sanjiv Das, George
Chacko and Michael Spalinski for stimulating interactions. I also thank
Cumrun Vafa and the string theory group for their kind hospitality.
\end{center}

\vskip 0.3cm
\noindent

\section*{Appendix}

\appendix

\sect{Lattice Formulation}

We do a more careful and rigorous treatment of the field theory for the
Treasury bonds. We first discretize the variables into a lattice of
discrete points. Let
$(t,x)\rightarrow (m\epsilon,na)$, where $\epsilon$ is an infinitesimal
time step and $a$ is an infinitesimal in the $x$ direction. Consider the
trapezoidal domain ${\cal T}$ given in Fig.(\ref{figa}) to be bounded by
integers  $m=m_0(={t_0\over \epsilon}),m=m_*(={t_*\over
\epsilon}), m\epsilon=na$ and $n=N(={T\over a})$. The
integers then take values in the 
lattice version of the trapezoidal domain, say ${\cal D}$ given by

\eq
{\cal D}=\{m=m_0,m_0+1,...m_*; n=m,m+1,....N\}
\qe 

The forward rates and quantum field yield on discretization 

\eqr
f(t,x)&\rightarrow& f(m\epsilon,na)\equiv f_{mn}\\
A(t,x)&\rightarrow& A(m\epsilon,na)\equiv A_{mn}   
\rqe

and similarly for $\alpha$ and $\sigma$.

From eqn.(\ref{dfA}) we have

\eq
f_{m+1n}=f_{mn}+\epsilon \alpha_{mn}+\epsilon \sigma_{mn}A_{mn}
\qe

Using finite differences to discretize derivatives, the generalization of
action $S_0$ in eqn.(\ref{s0}) is given, for $s=\sqrt{Na-m\epsilon}$ , by

\eqr
\label{slat}
S[A]&=&-{\epsilon\over
2}\sum_{m=m_0+1}^{m_*}\Big\{{a\over{(N+1)a-m\epsilon}}\sum_{n=m}^{N} A^2_{mn}+ 
{1\over{Na-m\epsilon}}{a\over{\mu^2}}\sum_{n=m}^{N-1}(A_{mn+1}-A_{mn})^2\Big\}\nonumber  \\
\\
\label{dA}
\int dA&=&\prod_{m=m_0+1}^{m_*}\sqrt{\epsilon\mu s\over{2\pi\sinh\mu s
}}\prod_{n=m}^N \sqrt{\epsilon\over{2\pi \mu^2
a}}\int_{-\infty}^{\+\infty} dA_{mn} 
\rqe

Note the functional integral over the field $A(t,x)$ has been reduced to
a {\it finite-dimensional} multiple integral over the $A_{mn}$ variables,
which in the case 
above consists of $(m_*-m_0)\{N-(m_0+m_*-1)/2\}$ independent variables; hence
all the techniques 
useful for evaluating finite dimensional integrals can be used for
performing the integration over $A_{mn}$.

To achieve the correct normalization, one in fact need not keep track of
all the tedious pre-factors in (\ref{dA}). Instead one simply redefines the
action by  

\eqr
e^{S[A]}&\to&e^{S[A]}/Z\\
\label{za1}
Z&=&\int dA e^{S[A]}
\rqe

All the pre-factors in (\ref{dA}) cancel out; and more importantly, the
expression $e^S/Z$ is correctly normalized to be interpreted as a
probability distribution, and hence can be used for Monte Carlo studies of
this theory. The action given in (\ref {slat}) is the starting point for
any simulations 
that are required of the model including the pricing of path
dependent derivatives; there are well known
numerical algorithms developed in physics for numerically studying quantum
fields \cite{he}. 

We explicitly solve for the case of $\mu\rightarrow 0$ to see how the
HJM-model emerges. For  $\mu\rightarrow 0$, the second term in the action
gives a product of $\delta$-functions and we have

\eqr
\label{da}
e^{S[A]}&=&e^{S_0}\prod_{m=m_0+1}^{m_*}\prod_{n=m}^{N-1}\delta(A_{mn+1}-A_{mn})
\\
S_0&=&-{\epsilon\over
2}\sum_{m=m_0+1}^{m_*}{a\over{(N+1)a-m\epsilon}}\sum_{n=m}^{N} A^2_{mn}
\rqe   

Consider evaluating a typical expression like $Z$ in (\ref{za}).  For each
$m$, there are N-m+1 integration variables $A_{mn}$; from 
eqn.(\ref{da}) we see that there are $N-m$ $\delta$-functions, leaving
only one variable, say $A_{mm}$ unrestricted.  For simplicity, we take
$\epsilon=a$; hence we have 

\eqr
Z&=&\prod_{m=m_0+1}^{m_*}\sqrt{\epsilon\over{2\pi}}\int dA_{mm} e^{S_0}\\
S_0&=&-{\epsilon\over2}\sum_{m=m_0+1}^{m_*}A^2_{mm}
\rqe

Defining $W(m)=A_{mm}$, we see from eqns. (\ref{P}) that we have recovered
the HJM-model. We can equivalently consider 

\eq
W(m)={1\over{N-m+1}}\sum_{n=m}^N A_{mn}
\qe

and we have

\eq
\lim_ {\mu\rightarrow 0} W(m)\to A_{mm}
\qe

Taking the continuum limit, we see that the field theory, in the $\mu
\rightarrow 0$ limit reduces to 

\eqr
\label{hjm}
S_0\to -{1\over 2}\int_{t_0}^{t_*}dtW^2(t)\\
W(t)={1\over{T-t}}\int_{t}^Tdx A(t,x)
\rqe    

For the general case of $\mu \ne 0$, from eqn.(\ref{slat}),taking the continuum limit of
$\epsilon \to 0,a\to 0$ we finally obtain 

\eqr
S[A]&=&-{1\over2}\int_{t_0}^{t_*}dt{1\over{T-t}}\int_t^T
dx\{A^2(t,x)+{1\over\mu^2}({\partial A(t,x)\over{\partial
x}})^2 \}\\
\int DA&=&\prod_{(t,x)\epsilon {\cal T}}\int dA \equiv \lim_{\epsilon \to
0,a\to 0}\prod_{mn}\int dA_{mn}\\ 
Z&=&\int DA e^{S[A]}
\rqe

\sect{Generating Functional Z[J]}

Since the generating functional $Z[J]$  has been of central importance in
studying the 
quantum field theory, for completeness we briefly discuss its derivation; all
these results are well-known in physics \cite{zj} and this derivation is
intended for readers from other disciplines.

Recall

\eqr  
Z[J]&=&{1\over Z}\int DA e^{S[A,J]}\\
S[A,J]&=& \int_{t_0}^{t_*}dt\int_t^TdxJ(t,x)A(t,x)+S[A]
\rqe

Since $S[A,J]$ is quadratic functional of the field $A(t,x)$, to perform the
functional integration over the field, all we need to
do is to find the specific configuration of $A(t,x)$, say $a(t,x)$ which
maximizes $S[A,J]$; due to our choice of  normalization  $Z[J]$
depends only on $a(t,x)$.      

Since there is no coupling in the time direction $t$, we study the
solution $a(t,x)$ separately for each t, and on the finite line interval
$t<x<T$.  We first study
the case for which the boundary values of the field $A(t,x)$ are 
fixed, that is consider $A(t,t)=p$ and $A(t,T)=p'$ to be held fixed; we
will later integrate over $p,p'$ as is required for the evaluation of $Z[J]$.
We henceforth suppress the time variable $t$ for notational convenience.

The 'classical' (deterministic) field configuration $a(t,x)\equiv a(x)$ is
defined by 

\eqr
\label{fe}
{{\delta S[a,J]}\over{\delta A(t,x)}}&=&0\\
\label{bc}
a(x=t)&=&p;a(x=T)=p'
\rqe

Doing a change of variables $A(t,x)=B(t,x)+a(t,x)$ and a functional
Taylors expansion we have, from eqn.(\ref{fe})

\eqr
\label{cls}
S[A+a,J]=S_{cl}[a,J]+ {\tilde S}[B]
\rqe

where due to boundary conditions given in eqn.(\ref{bc}) ${\tilde S}[B]$ is
independent of $p,p',J$.  The functional integral over the $B(t,x)$
variables gives only an overall constant which we can ignore and hence we
have   

\eq
\label{zq}
Z[J]={1\over Z}\int_{-\infty}^{+\infty}dpdp'e^{S_{cl}[a,J]}
\qe

We now determine $a(x)$; from (\ref{fe}) we have

\eqr
\label{ode}
{1\over{\mu^2}}{{\partial^2a(x)}\over{\partial x^2}}
-a(x)+(T-t)J(x)=0 \\
a(t)=p,a(T)=p'; t<x<T
\rqe

Since eqn.(\ref{ode}) is a linear, the solution for $a(x)$ is given by a
sum of the 
solutions of the homogeneous and inhomogeneous equations; it can be verified
that, for $ \beta =T-t$, we have

\eqr
\label{soln}
a(x)&=&{\beta\over{sinh(\mu\beta)}}[a_H(x)+a_{IH}(x)]\\
\rqe

with the homogeneous solution given by

\eqr
a_H(x)&=&psinh\mu(T-x)+p'sinh\mu(x-t)
\rqe

and the inhomogeneous solution given by

\eqr
a_{IH}(x)&=&\mu
\int_t^T dx'[\theta(x-x')sinh\mu(T-x)sinh\mu(x'-t)\nonumber\\
   &+&\theta(x'-x)sinh\mu(T-x')sinh\mu(x-t)]J(x')
\rqe    

The 'classical' action is given by

\eqr
\label{s1}
S_{cl}[a,J]=S_1[p,p';J]+S_2[J]
\rqe

with

\eqr
\label{s2}
S_1[p,p';J]&=&-{1\over{2\mu\beta
 sinh(\mu\beta)}}\Big\{cosh(\mu\beta)(p^2+{p'}^2) -2pp'\Big\} 
 +{1\over{sinh(\mu\beta)}}[pP+p'Q]\nonumber\\
\\
P&=&\int_t^Tdxsinh\mu(x-t)J(x),  Q=\int_t^Tdx sinh\mu(T-x)J(x)
\rqe

and 

\eq
\label{s3}
S_2[J]={\mu\beta\over{sinh(\mu\beta)}}\int_t^Tdxdx'\theta(x-x')sinh\mu(T-x)
sinh\mu(x'-t)J(x)J(x')    
\qe

Performing the Gaussian integrations over $p,p'$ and restoring the
time variable $t$ yields

\eqr
\label{zf}
Z[J]&=&{1\over Z}e^{S_2[J]}\int dpdp'e^{S_1[p,p';J]}\\
    &=&\exp
{1\over2}\int_{t_0}^{t_*}dt\int_t^Tdxdx'J(t,x)D(x,x';t,T)J(t,x') 
\rqe

where, from eqns.(\ref{s1}), (\ref{s2}) and (\ref{s3}) we have

\eqr
\label{pro}
D(x,x';t,T)&=&{{\mu\beta}\over{
\sinh^3(\mu\beta)}}\Big [\sinh\mu(T-x)\sinh\mu(x'-t)\{1+\sinh^2(\mu
\beta)\theta(x-x')\}   
\nonumber\\    
&+&\sinh\mu(T-x')\sinh\mu(x-t)\{1+\sinh^2(\mu\beta)\theta(x'-x)\}\nonumber \\
    &+&\cosh(\mu\beta)\{\sinh\mu(x-t)\sinh\mu(x'-t) \nonumber\\
            &+&\sinh\mu(T-x)\sinh\mu(T-x')\} \Big ] 
\rqe

\end{document}